\documentclass[%
 reprint,
 amsmath,amssymb,
 aps,
]{revtex4-2}

\usepackage{graphicx}
\usepackage{dcolumn}
\usepackage{bm}
\usepackage{color}

\def\s{{\sigma}}
\def\e{{\varepsilon}}

\def\0{{ {\bm 0} }}

\def\w{{\omega}}
\def\a{{\alpha}}

\allowdisplaybreaks[4]

\makeindex

\begin{document}

\title{
Emergence of $d \pm ip$-wave superconducting state at the edge of $d$-wave superconductors
mediated by Andreev-bound-state-driven ferromagnetic fluctuations
}

\author{Shun Matsubara}
\author{Hiroshi Kontani}
\affiliation{Department of Physics, Nagoya University, 
Nagoya 464-8602, Japan}

\date{\today}

\begin{abstract}
We propose a mechanism of spin-triplet superconductivity
at the edge of $d$-wave superconductors.
Recent theoretical research in $d$-wave superconductors
predicted that strong ferromagnetic (FM) fluctuations
are induced by large density of states due to edge Andreev bound states (ABS).
Here, we construct the linearized gap equation for the edge-induced superconductivity,
and perform a numerical study based on a large cluster Hubbard model
with bulk $d$-wave superconducting (SC) gap.
We find that ABS-induced strong FM fluctuations
mediate the $d \pm ip$-wave SC state,
in which the time-reversal symmetry is broken.
The edge-induced $p$-wave transition temperature $T_{cp}$
is slightly lower than the bulk $d$-wave one $T_{cd}$,
and the Majorana bound state may be created at the endpoint of the edge.
\end{abstract}


\maketitle

\section{Introduction}

In cuprate high-$T_c$ superconductors,
spin fluctuations induce various kind of interesting phenomena.
For example, $d$-wave superconductivity is mediated by the antiferromagnetic (AFM) fluctuations
\cite{Bickers,Monthoux_FLEX,Koikegami_FLEX,Takimoto_FLEX,Dahm_FLEX,Manske_FLEX}.
Non-Fermi-liquid transport phenomena
such as $T$-linear resistivity,
Curie-Weiss behavior of the Hall coefficient,
and the modified Kohler rule between the magnetoresistance
and Hall angle ($\Delta\rho/\rho_0\propto(\sigma_{xy}/\sigma_{xx})^2)$)
are understood as the effects of strong AFM fluctuations on the Fermi liquid state
\cite{Moriya,Moriya-2,Pines,Kontani-rev,Kontani-Hall,Kontani-thermoelectric,Kontani-Nernst-magresi}.
Moreover,
recently discovered
axial and uniform charge density wave (CDW) 
\cite{CDW_Ghiringhelli,CDW_Chang,CDW_Fujita,uniform_CDW_Matsuda}
has been theoretically understood as
the spin-fluctuation-driven CDW
due to Aslamazov-Larkin vertex correction mechanism.
\cite{Chubukov_CDW,Kivelson_CDW,Sachdev_CDW,Onari-CDW,Yamakawa-CDW,Kawaguchi-CDW}.

In addition, 
by introducing real-space structures such as surfaces and impurities,
interesting non-trivial critical phenomena emerge in correlated electron systems.
In cuprate superconductors,
non-magnetic impurities enhance the spin fluctuations around them
\cite{Alloul99-2,Ishida96,Alloul94,Alloul00,Alloul00-2,Alloul99,Bulut00,Ohashi_imp_RPA,Kontani-imp}.
In the two-dimensional Hubbard model with the ($1,1$) edge,
the ferromagnetic (FM) fluctuations develop along the edge
\cite{Matsubara-edge}.
These phenomena are caused by the Friedel oscillation
in the local density of states (LDOS)
since the large LDOS sites near the real-space structure
drive the system toward the magnetic criticality.

In contrast, in the superconducting (SC) states,
studies of the effects of real-space structures
on the electron correlation were limited until recently.
Recently, several interesting impurity-induced
\cite{imp_Harter,suf_Andersen}
and surface-induced
\cite{matsubara_abs_fm}
critical phenomena have been analyzed theoretically.
The key ingredient is the edge-induced Andreev bound states (ABS)
in the $d$-wave superconductors
\cite{Hu-ZBCP,Tanaka-ZBCP,Kashiwaya-junction,Matsumoto-Shiba-ABS,Nagato,Kashiwaya-ZBCP},
which is observed in the STM experiment as the zero-bias conductance peak
\cite{Kashiwaya-ZBCP-2,Iguchi,Wei-ZBCP,Geek-ZBCP}.
In a previous paper
\cite{matsubara_abs_fm},
the present authors revealed that
the huge edge DOS due to the ABS triggers very strong FM fluctuations
around the ($1,1$) edge,
by carrying out site-dependent random-phase approximation (RPA)
and modified fluctuation-exchange (FLEX) approximation.
In this case, the strong FM fluctuations may induce exotic phenomena such as the triplet superconductivity
\cite{triplet_FM_fluc_Fay,triplet_FM_fluc_Monthoux,triplet_FM_fluc_Wang,triplet_FM_fluc_Roussev,triplet_FM_fluc_Fujimoto}.

As well-known,
the emergence of surface or interface induced SC state that is not realized in the bulk
has been studied very actively.
Near the ($1,1$) edge of the $d_{x^2-y^2}$-wave superconductor, 
the $s$-wave superconductivity can emerge by using the ABS,
and an $d \pm is$-wave SC state is realized
\cite{Matsumoto-Shiba-I,Matsumoto-Shiba-II,Matsumoto-Shiba-III,Tanuma_dpis,zbcp_split_1,zbcp_split_2,zbcp_split_3}.
In this case, time-reversal symmetry is broken
and the zero-bias conductance peak splits.
In addition, the edge current flows along the edge.
This emergence of time-reversal breaking superconductivity at the domain wall
is also discussed with regards to the polycrystalline
${\rm Y}{\rm Ba}_2{\rm Cu}_3{\rm O}_{7-x}$ (YBCO) \cite{Sigrist-Kuboki-TB,Kuboki-TB_t-J,Kuboki-GL_jpsj}
and twined iron-based superconductor FeSe in the nematic phase \cite{Watashige-FeSe-TB}.
However, the site-dependence of pairing interaction has not been taken into consideration,
although FM fluctuations are strongly enhanced near the edge of the Hubbard model.
Recently, the emergence of the fractional vortices and supercurrent near the ($1,1$) edge
is proposed \cite{supercurrent_1,supercurrent_2}.
In this case, the ABS is shifted to the finite energy and the time-reversal symmetry is broken.

In this paper, we theoretically predict the emergence of 
the triplet superconductivity near the ($1,1$) edge of the $d$-wave superconductors.
The origin of the triplet gap is the strong FM fluctuations triggered by the ABS
due to the sign-change in the $d$-wave SC gap.
We first develop the linearized gap equation
for the edge-superconductivity, and apply it to
a two-dimensional cluster Hubbard model
with the ($1,1$) edge in the bulk $d$-wave SC state.
The site-dependent pairing interaction is obtained
based on the microscopic calculation by the RPA or $GV^I$-FLEX
\cite{matsubara_abs_fm}.
We reveal that the phase difference between the edge triplet gap
and the bulk $d$-wave gap is $\pi/2$ in the ${\bm k}$-space.
That is, exotic edge-induced
$d \pm ip$-wave SC state is expected to be realized
at $T=T_{cp}$, which is slightly lower than
the bulk $d$-wave transition temperature $T_{cd}$.
The present study may offer an interesting platform of realizing exotic SC states.

\section{Theoretical method of  triplet gap equation}
\begin{figure}[t]
\includegraphics[width=0.9\linewidth]{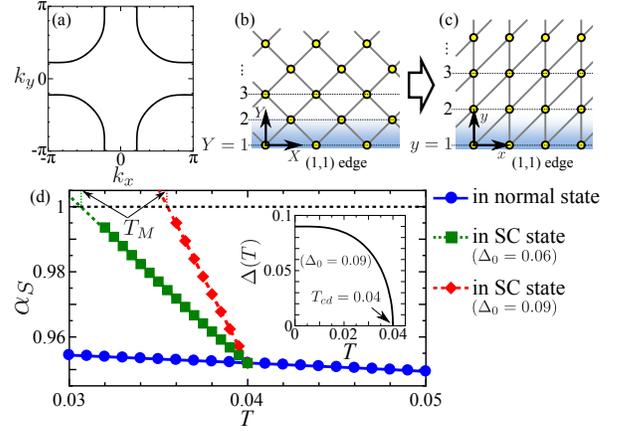}
\caption{(color online)
(a) Fermi surface in the bulk YBCO TB model at filling $n=0.95$.
(b) Square lattice with ($1,1$) edge.
(c) One-site unit cell square lattice with ($1,1$) edge.
To simplify the calculation, we actually use the square lattice shown in (c) instead of (b).
(d) $T$-dependence of $\alpha_S$ in the RPA.
The inset shows the $T$-dependence of the bulk $d$-wave gap given in \eqref{eq:d-wave-t-dep}.
We set the transition temperature of the $d$-wave superconductivity as $T_{cd}=0.04$.
At $T=T_{M}$, $\alpha_S$ reaches unity.
} 
\label{fig:fig1}
\end{figure}

To study the edge-induced triplet superconductivity,
we construct a two-dimensional square lattice Hubbard model
with the ($1,1$) edge in the bulk $d$-wave SC state:
\begin{eqnarray}
\mathcal{H}&=&\sum_{i,j,\s}t_{i,j}c_{i\s}^\dagger c_{j\s}
+U\sum_{i}n_{i\uparrow}n_{i\downarrow}
\nonumber\\
& &+\sum_{i,j}
\left(
\Delta_{i,j}^{\uparrow\downarrow}
c_{i\uparrow}^\dagger c_{j\downarrow}^\dagger
+
h.c.
\right),
\label{eqn:Hamiltonian}
\end{eqnarray}
where $t_{i,j}$ is the hopping integral between sites $i$ and $j$.
We set the nearest, next nearest, and third-nearest hopping integrals as $(t,t',t'')=(-1,1/6,-1/5)$,
which correspond to the YBCO TB model.
$c_{i\sigma}^\dag$ and $c_{i\sigma}$
are creation and annihilation operators of an electron with spin $\sigma$, respectively.
$U$ is the on-site Coulomb interaction, and
$\Delta_{i,j}^{\uparrow\downarrow}=-\Delta_{i,j}^{\uparrow\downarrow}\equiv\Delta_{i,j}$
is the bulk $d$-wave SC gap.
Figure \ref{fig:fig1} (a) shows the Fermi surface of the periodic tight-binding (TB) model at filling $n=0.95$.
In this model, the AFM fluctuations develop in the bulk due to the nesting $Q\approx(\pi,\pi)$.
Fig. \ref{fig:fig1} (b) shows the original square lattice with the ($1,1$) edge.
If we analyze the original square lattice along the $X$- and $Y$-axis,
there are two sites in an unit cell, and it makes the analysis complicated.
For convenience,
we analyze an equivalent ($1,1$) edge model with the one-site unit cell structure
shown in Fig. \ref{fig:fig1} (c).
$y=1$ corresponds to the ($1,1$) edge layer.
This model is periodic along the $x$ direction,
whereas the translational symmetry along $y$ direction is violated.
Thus, we perform following analysis in $(k_x,y,y')$-representation
obtained by the Fourier transformation only on the $x$ direction.
Here, we represent the Fourier transformation of the first term of \eqref{eqn:Hamiltonian} as follows:
\begin{eqnarray}
H^0=
\sum_{k_x,y,y',\s}H_{y,y'}^0(k_x)c_{k_x,y,\s}^\dagger c_{k_x,y',\s}.
\label{eqn:tight-binding}
\end{eqnarray}

Next, we assume that $\Delta_{i,j}$
is real and nonzero only between the nearest neighbor sites,
and set it as
$
\Delta_{i,j}=
\Delta/2
(
\delta_{x,x'+1}
\delta_{y,y'+1}
+
\delta_{x,x'-1}
\delta_{y,y'-1}
-
\delta_{x,x'}
\delta_{y,y'+1}
-
\delta_{x,x'}
\delta_{y,y'-1}
)
$.
By performing the Fourier transformation on $x$ direction,
we obtain its ($k_x,y,y'$)-representation as
\begin{eqnarray}
& &
\Delta_{y,y'}(k_x,T)
\nonumber \\
& &
=\Delta(T)
\left\{
\frac{e^{-ik_x}-1}{2}\delta_{y,y'+1}
+\frac{e^{ik_x}-1}{2}\delta_{y,y'-1}
\right\},
\label{eqn:d-gap}
\end{eqnarray}
\begin{eqnarray}
\Delta(T)
=\Delta_0
\tanh\left(1.74\sqrt{\frac{T_{cd}}{T}-1}\right),
\label{eq:d-wave-t-dep}
\end{eqnarray}
where $\Delta(T)$ is the temperature-dependent $d$-wave gap
and $\Delta_0\equiv\Delta(T=0)$.
Note that $\Delta(\bm{k},T)=\Delta(T)(\cos{k_x}-\cos{k_y})$ in a bulk $d$-wave superconductor.
$T_{cd}$ is the transition temperature of the $d$-wave superconductivity.
Here, we confirm the relations of the bulk $d$-wave gap.
Due to the anticommutation relation of the fermion,
the SC gap satisfies
\begin{align}
\Delta_{y,y'}(k_x)\equiv
\Delta_{y,y'}^{\uparrow\downarrow}(k_x)=-\Delta_{y',y}^{\downarrow\uparrow}(-k_x).
\label{eq:singlet-anticom}
\end{align}
The definition of the singlet gap is
\begin{align}
\Delta^{\uparrow\downarrow}_{y,y'}(k_x)=-\Delta^{\downarrow\uparrow}_{y,y'}(k_x).
\label{eq:singlet-def}
\end{align}
By using \eqref{eq:singlet-anticom} and \eqref{eq:singlet-def},
the singlet gap satisfies
\begin{align}
\Delta^{\uparrow\downarrow}_{y,y'}(k_x)=\Delta^{\uparrow\downarrow}_{y',y}(-k_x).
\label{eq:singlet_even-parity}
\end{align}
Since we set $\Delta_{i,j}$ without the loss of generality,
the present real $d$-wave gap given by \eqref{eqn:d-gap} satisfies
\begin{align}
{\Delta_{y,y'}^{\uparrow\downarrow}}^*(-k_x)=\Delta_{y,y'}^{\uparrow\downarrow}(k_x).
\label{eq:singlet_delta_deltadag}
\end{align}
Hereafter, we introduce the $N_y \times N_y$ matrix representations of
the $d$-wave gap function $\hat{\Delta}(k_x)$,
which is defined as
$\{\hat{\Delta}(k_x)\}_{y,y'}$=$\Delta_{y,y'}(k_x)$.

We also define $N_y\times N_y$ Green functions in the $d$-wave SC state
$\hat G$, $\hat F$, and $\hat F^\dag$ as follows:
\begin{eqnarray}
& &
  \left(
    \begin{array}{cc}
     \hat{G}(k_x,\e_n) & \hat{F}(k_x,\e_n) \\
     \hat{F}^{\dag}(k_x,\e_n) & -\hat{G}(k_x,-\e_n)  
    \end{array}
  \right)
\nonumber\\
& &
=
  \left(
    \begin{array}{cc}
     \e_n\hat{1}-\hat{H}^0(k_x) & -{\hat{\Delta}}(k_x) \\
     -{\hat{\Delta}}(k_x) & \e_n\hat{1}+\hat{H}^0(k_x)
    \end{array}
  \right)^{-1},
\label{eq:sc-4}
\end{eqnarray}
where  $\e_n=(2n+1)\pi iT$ is the fermion Matsubara frequency.
$\hat{F}$ and $\hat{F}^\dag$ are anomalous Green functions, which are finite only in the bulk $d$-wave SC state.
Since the $d$-wave gap satisfies \eqref{eq:singlet-def},
the anomalous Green function $\hat{F}$ satisfies the relation
\begin{align}
\hat{F}^{\uparrow\downarrow}=-\hat{F}^{\downarrow\uparrow}
\equiv
\hat{F}.
\label{eq:singlet-f}
\end{align}

In this model,
we can obtain the enhancement in the FM fluctuations at the edge
by the RPA or $GV^I$-FLEX approximation \cite{matsubara_abs_fm}.
In these analyses, we define the irreducible susceptibilities as follows:
\begin{eqnarray}
\chi^0_{y,y'}({q}_x,\w_l) &=&-T\sum_{{k}_x,n}
G_{y,y'}({q}_x+{k}_x,\w_l+\e_n)
\nonumber\\
&&\times G_{y',y}({k}_x,\e_n) ,
\label{eqn:chi0}
\end{eqnarray}
\begin{eqnarray}
\varphi^0_{y,y'}(q_x,\omega_l)
&=&
-T\sum_{k_x,n}
F_{y,y'}(q_x+k_x,\omega_l+\e_n)
\nonumber\\
&&\times
F_{y',y}^{\dag}(k_x,\e_n) ,
\label{eqn:phi0}
\end{eqnarray}
where  $\omega_l=2l\pi iT$ is the boson Matsubara frequency.
$\hat{\varphi}^0$ is finite only in the SC state.
The site-dependent spin susceptibility $\hat{\chi}^s$ is calculated
using $\hat{\chi}^0$ and $\hat{\varphi}^0$ as
\begin{eqnarray}
\hat{\chi}^s(q_x,\omega_l)
&=&
\hat{\Phi}(q_x,\omega_l)
\left\{
\hat{1}-U
\hat{\Phi}(q_x,\omega_l)
\right\}^{-1}.
\label{eqn:chis}
\end{eqnarray}
\begin{eqnarray}
&&
\hat{\Phi}(q_x,\omega_l)
=
\hat{\chi}^0(q_x,\omega_l)+\hat{\varphi}^0(q_x,\omega_l),
\label{eqn:phisc}
\end{eqnarray}
The spin Stoner factor, $\a_S$, is defined as the largest eigenvalue of
$U\hat{\Phi}(q_x,\omega_l)$ at $\w_l=0$.
It represents the spin fluctuation strength,
and the magnetic order is realized when $\a_S\ge1$.
Fig. \ref{fig:fig1} (d) shows the $T$-dependence of the Stoner factor $\alpha_S$ in the RPA.
The inset shows the $T$-dependence of the bulk $d$-wave gap given by \eqref{eq:d-wave-t-dep}.
In the $d$-wave SC state, $\alpha_S$ drastically increases as $T$ decreases
due to the development of the ABS.
In this case, the static spin susceptibility along the ($1,1$) edge layer $\chi^s_{1,1}(q_x,0)$ has large peak at $q_x=0$.
This edge FM correlation is consistent with the bulk AFM correlation.
At $T=T_M$, $\alpha_S$ reaches unity and edge FM order is realized.

Next, we analyze the edge-induced triplet superconductivity in the presence of the bulk $d$-wave SC gap.
Here, we represent the triplet SC gap in ($k_x,y,y'$)-representation as
$\phi^{\uparrow\downarrow}_{y,y'}(k_x)$.
In this study, we do not consider the spin orbit interaction.
Then we can set the d-vector as $\hat{\bm d}(k_x)=(0,0,\hat{\phi}(k_x))$ without losing generality.
In this case,
we consider only $\phi^{\uparrow\downarrow}_{y,y'}(k_x)$ and $\phi^{\downarrow\uparrow}_{y,y'}(k_x)$.
Due to the anticommutation relation of the fermion,
the SC gap satisfies
\begin{align}
\phi_{y,y'}(k_x)
\equiv
\phi_{y,y'}^{\uparrow\downarrow}(k_x)=-\phi_{y',y}^{\downarrow\uparrow}(-k_x).
\label{eq:triplet_anti_com}
\end{align}
The definition of the triplet gap is
\begin{align}
\phi^{\uparrow\downarrow}_{y,y'}(k_x)=\phi^{\downarrow\uparrow}_{y,y'}(k_x).
\label{eq:triplet-def}
\end{align}
From, \eqref{eq:triplet_anti_com} and \eqref{eq:triplet-def},
the triplet gap follows
\begin{align}
\phi^{\uparrow\downarrow}_{y,y'}(k_x)=-\phi^{\uparrow\downarrow}_{y',y}(-k_x).
\label{eq:triplet_odd-parity}
\end{align}
Here, we introduce $N_y\times N_y$ matrix representation $\hat{\phi}(k_x)$, 
which is defined as $\{\hat{\phi}(k_x)\}_{y,y'}$=$\phi_{y,y'}(k_x)$.
To decide the edge-induced SC state,
we must obtain the phase difference between the bulk $d$-wave gap and the edge triplet gap.
Although we can use the Bogoliubov-de Gennes (BdG) equation,
we have to perform heavy self-consistent calculation at various temperatures.
To make the theoretical analysis much more efficient,
we develop the linearized gap equation for the edge triplet superconductivity,
by linearizing the BdG equation only for $\hat{\phi}$ and $\hat{\phi}^\dag$.
We set the eigenvalue of the linearized equation as $\lambda$.
When $\lambda\ge1$,
the triplet superconductivity emerges and coexists with the bulk $d$-wave superconductivity.
In this method, by just performing the diagonalization,
we can address the emergence of triplet superconductivity by the temperature-dependence of the eigenvalue .
We show the details of the derivation of the linearized equation in Appendix A and B.
We use the relation \eqref{eq:singlet-f} and \eqref{eq:triplet-def} in the derivation of the linearized gap equation,
and it is given as
\begin{subequations}
\begin{eqnarray}
& &\lambda
\phi_{y,y'}(k_x)
\nonumber\\
&=&
-
T
\sum_{k_x',Y,Y',n}
V_{y,y'}(k_x-k_x',\e_n-\e_0)
\nonumber\\
& &
\times
\left\{
G_{y,Y}(k_x',\e_n)
\phi_{Y,Y'}(k_x')
G_{y',Y'}(-k_x',-\e_n)
\right.
\nonumber\\
& &
-
\left.
F_{y,Y}(k_x',\e_n)
{\phi^{\dag}_{Y,Y'}}(k_x')
F_{Y',y'}(k_x',\e_n)
\right\},
\label{eq:trip_gap_eq1}
\end{eqnarray}
\begin{eqnarray}
& &\lambda
{\phi^\dag_{y,y'}}(k_x)
\nonumber\\
&=&
-
T
\sum_{k_x',Y,Y',n}
V_{y,y'}(k_x'-k_x,\e_n-\e_0)
\nonumber\\
& &
\times
\left\{
G_{Y,y}(-k_x',-\e_n)
{\phi^\dag_{Y,Y'}}(k_x')
G_{Y',y'}(k_x',\e_n)
\right.
\nonumber\\
& &
-
\left.
F_{y,Y}^\dag(k_x',\e_n)
\phi_{Y,Y'}(k_x')
F_{Y',y'}^\dag(k_x',\e_n)
\right\},
\label{eq:trip_gap_eq2}
\end{eqnarray}
\label{eq:trip_gap_eq3}
\end{subequations}
\begin{eqnarray}
\hat{V}({q}_x,\w_l)
=U^2\left(-\frac12\hat{\chi}^s({q}_x)
-\frac12\hat{\chi}^c({q}_x)\right)
C(\w_l,\w_d),
\nonumber\\
\label{eq:trip_gap_int}
\end{eqnarray}
where $\hat{V}({q}_x,\w_l)$ is the site-dependent pairing interaction for triplet superconductivity.
$\hat{\chi}^{s(c)}(q_x)$ is the static spin (charge) susceptibility in the $d$-wave SC state
obtained by the RPA or $GV^I$-FLEX approximation. 
Here, $\w_l=2l\pi iT$ is the boson Matsubara frequency.
$C(\w_l,\w_d)={\omega_d^2}/\left({|\w_l|^2+\w_d^2}\right)$ is a cut off function,
where $\w_d$ is the cutoff energy, and we set $\w_d=0.5$. 
We then solve the gap equation \eqref{eq:trip_gap_eq3} under the restriction \eqref{eq:triplet_odd-parity}.
Note that the first and second terms of the gap equation have different sign
due to the relation \eqref{eq:singlet-f}.
This fact greatly affects the phase difference
between the bulk gap function and the edge one.

\begin{figure}[h]
\includegraphics[width=0.9\linewidth]{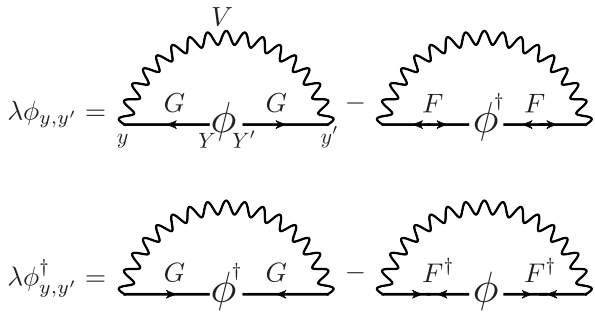}
\caption{(color online)
Diagram of the linearized triplet SC gap equation in the presence of the bulk $d$-wave SC gap.
The undulating lines are pairing interactions of the triplet superconductivity.
The line with a single arrow represents the Green function $\hat{G}$
and the line with double arrows represents anomalous Green functions
$\hat{F}$ and $\hat{F}^{\dag}$.
}
\label{fig:fig2}
\end{figure}

Figure \ref{fig:fig2} is the diagrammatic expression of the gap equation \eqref{eq:trip_gap_eq3}.
The undulating lines are pairing interactions $\hat{V}$.
The diagrams with $GG$ correspond to the conventional gap equation in the normal state.
The diagrams with $FF$ are newly added
to describe the effect of the bulk $d$-wave SC gap on the edge superconductivity.
Since $\hat{\phi}$ and $\hat{\phi}^{\dag}$ are mixed in the present gap equation developed in Eq. \eqref{eq:trip_gap_eq3},
the phase of $\hat {\phi}$ is uniquely determined.
From the view point of the Ginzburg-Landau (GL) theory,
the diagrams with $GG$ and those with $FF$ in Fig. \ref{fig:fig2} respectively
give rise to
the fourth-order term $|\Delta|^2|\phi|^2$ or ${\rm Re}\{\Delta^2{\phi^{*}}^2\}$ in the free energy.
The latter GL term determines
the phase difference between $\hat{\Delta}$ and $\hat{\phi}$.

\section{numerical result of triplet gap equation}
In this section, we analyze the linearized triplet gap equation \eqref{eq:trip_gap_eq3}.
$k_x$-mesh is $N_x=64$,
site number along $y$-direction is $N_y=64$,
the number of Matsubara frequencies is 1024.
The transition temperature of the bulk $d$-wave superconductivity is $T_{cd}=0.04$.
The Coulomb interaction is $U=2.25$ in the RPA, and $U=2.65$ in the $GV^I$-FLEX.
Here, the unit of energy is $|t|$, which corresponds to 
$\sim 0.4$eV in cuprate superconductors.
In addition, we define $\Delta_{\rm max}$ as the maximum value of the $d$-wave gap on the Fermi surface.
In the present model, $\Delta_{\rm max}=1.76\Delta_0$ for $n=0.95$.
Experimentally, $4<2\Delta_{\rm max}/T_{cd}<10$ in YBCO
\cite{cuprate_coherence_1,cuprate_coherence_2}.
Therefore, in the RPA, we set $\Delta_0=0.06$ or 0.09, which corresponds to $\Delta_{\rm max}=5.28$ or 7.92 for $T_{cd}=0.04$.

\subsection{$d \pm ip$-wave SC state}
\begin{figure}[h]
\includegraphics[width=0.7\linewidth]{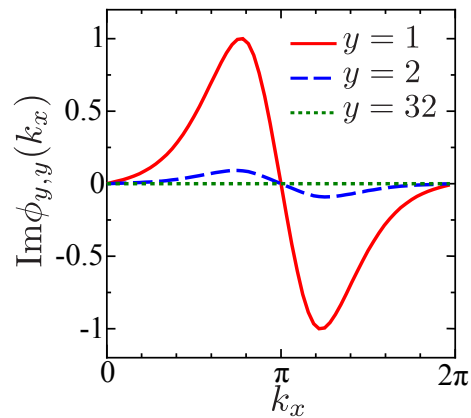}
\caption {(color online)
$k_x$-dependence of obtained $p_x$-wave SC gap $\phi_{y,y}(k_x)$
for $\Delta_0=0.09$ at $T=0.0375$.
The pairing interaction is calculated by the RPA.
$y=1$ and $y=32$ correspond to the edge and bulk, respectively.
We normalize the gap as $\underset{k_x,y}{\rm max}|\phi_{y,y}(k_x)|=1$.
}
\label{fig:fig3}
\end{figure}

First, we analyze the linearized triplet gap equation for the pairing interaction calculated by the RPA.
Figure \ref{fig:fig3} shows $k_x$-dependence of the obtained triplet gap in the same layer $y$.
This is the $p_x$-wave gap with a node at $k_x=0$.
It can emerge at the edge
because there are finite LDOS and large triplet pairing interactions due to the ABS.

Next, we discuss the phase difference between the $d$- and $p$-wave gap.
The triplet SC gap in the real space $\phi_{x,y,y'}$ is represented
by the Fourier transformation on the $x$-direction of $\phi_{y,y'}(k_x)$. 
By using \eqref{eq:triplet_odd-parity},
we obtain
\begin{align}
\phi_{x,y,y'}
=
-\left\{\sum_{k_x}\phi_{y,y'}^{\dag}(k_x)e^{ik_xx}\right\}^*.
\label{eq:fourier}
\end{align}
The relation
holds for the general triplet SC gap.
On the other hand,
the obtained $p$-wave gap satisfies
\begin{align} 
\phi_{y,y'}(k_x)=-\phi_{y,y'}^{\dag}(k_x),
\label{eq:condition_obtained-p-wave}
\end{align}
in the present numerical study.
Therefore, the obtained $p$-wave gap is a real function in real space $\phi_{x,y,y'}=\phi_{x,y,y'}^*$.
In this case, the phase difference is $\pm \pi/2$ in the $\bm{k}$-space, and this is the $d \pm ip$-wave SC state.
We find that the edge $d \pm ip$-wave SC state is stabilized
by the coexistence of the bulk $d$-wave superconductivity and the edge-induced triplet superconductivity.

The reason of this phase difference $\pm\pi/2$ is understood
by evaluating the contribution from the second term of \eqref{eq:trip_gap_eq3}.
Since the triplet pairing interaction $V_{y,y'}(k_x-k_x',\e_n-\e_0)$ has large value only at the edge ($y=1$) and $\Delta_{i,j}$ is real function,
we can approximately evaluate the contribution to $\phi_{1,1}(k_x)$ from second term of \eqref{eq:trip_gap_eq1}
by setting $Y=Y'=1$,
\begin{align}
&{\rm second\ term\ of\ \eqref{eq:trip_gap_eq1}}
\nonumber\\
&\approx
-
T\sum_{k_x',n}
|V_{1,1}(k_x-k_x',\e_n-\e_0)|
|{F_{1,1}(k_x',\e_n)}|^2
\phi^*_{1,1}(k_x').
\end{align}
Here, $V_{y,y'}(k_x-k_x',\e_n-\e_0)$ has a large peak at $k_x=k_x'$.
Therefore, the triplet superconductivity is stabilized
when $\phi^*_{1,1}(k_x)=\phi^\dag_{1,1}(k_x)=-\phi_{1,1}(k_x)$,
and it is actually confirmed by numerical calculation.

In the $d \pm ip$-wave SC state, the time-reversal (TR) symmetry is broken.
To verify it, we apply the time-reversal operator $\Theta=-i\sigma^yK$
to the present gap functions.
\begin{align}
{\Delta}^{\uparrow\downarrow}_{y,y'}(k_x)+{\phi}^{\uparrow\downarrow}_{y,y'}(k_x)
\xrightarrow[{\rm TR}]{}
-{{\Delta}^{\downarrow\uparrow}_{y,y'}}^*(-k_x)-{{\phi}^{\downarrow\uparrow}_{y,y'}}^*(-k_x).
\label{eq:tr-operation}
\end{align}
By using the conditions
\eqref{eq:singlet-def}, \eqref{eq:singlet_delta_deltadag},
\eqref{eq:triplet-def}, and \eqref{eq:condition_obtained-p-wave},
we confirm that the $d+ip$-wave gap changes to the $d-ip$-wave gap.
In Appendix C, we calculate the LDOS in the $d \pm ip$-wave SC state.
The LDOS for up spin electrons and that for down spin electrons are separated
since the time-reversal symmetry is broken in the $d \pm ip$-wave SC state.

\subsection{Temperature-dependence of $\lambda$}
\begin{figure}[h]
\includegraphics[width=0.7\linewidth]{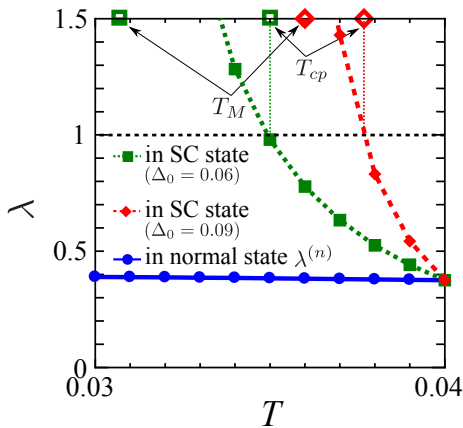}
\caption{(color online)
$T$-dependence of $\lambda$ for the pairing interaction by the RPA.
The red and green line represent $\lambda$ for $\Delta_0=0.06$ and
$0.09$, respectively.
The blue line shows $\lambda^{(n)}$ in the normal state ($\Delta_0=0$).
Below $T_{cp}$, the $p$-wave superconductivity emerges.
At $T=T_{M}$, $\alpha_S$ reaches unity in the RPA.
}
\label{fig:fig4}
\end{figure}

Next, we examine the $T$-dependence of the eigenvalue of the edge $p$-wave superconductivity.
We denote the eigenvalue
in the $d$-wave superconductivity and normal state as $\lambda$ and $\lambda^{(n)}$, respectively.
Figure \ref{fig:fig4} shows the $T$-dependence of the eigenvalue based on the RPA.
$\lambda^{(n)}$ hardly increases and does not reach unity.
On the other hand,
$\lambda$ increases drastically as $T$ decreases and exceeds unity
below $T_{cp} \lesssim T_{cd}$.
At these temperatures, the $d \pm ip$-wave SC state is realized.
Note that the edge FM order is realized at $T_{M} \lesssim T_{cp}$.
For $\Delta_0=0.09$ ($2\Delta_{\rm max}/T_{cd}=7.92$),
the increase in $\lambda$ is more drastic than that for $\Delta_0=0.06$ ($2\Delta_{\rm max}/T_{cd}=5.28$)
due to the stronger development of the FM fluctuations as shown in the Fig. \ref{fig:fig1}(d).

\begin{figure}[h]
\includegraphics[width=0.7\linewidth]{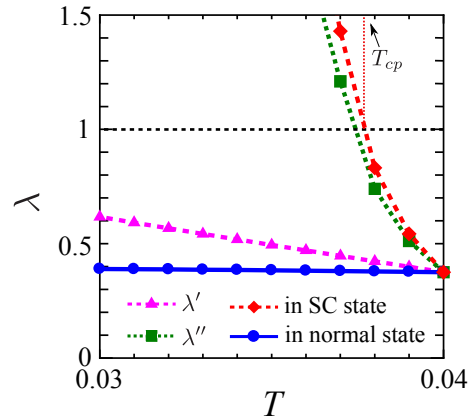}
\caption{(color online)
$T$-dependence of $\lambda'$ and $\lambda''$.
We set $\Delta_0=0.09$
The red doted line and blue solid line represent
$\lambda$ and $\lambda^{(n)}$ in the $d$-wave SC state and normal state, respectively.
}
\label{fig:fig5}
\end{figure}

To examine the effect of the FM fluctuations on the increase in $\lambda$,
we analyze two types of gap equations, (i) and (ii),
from which the effect of the $d$-wave gap is partially subtracted.
In (i), we use the pairing interaction in the normal state
$\hat{V}_{\rm normal}$ instead of $\hat{V}$ in the $d$-wave SC state,
and denote the eigenvalue as $\lambda'$.
In (ii), we replace the Green functions $\hat{G}$, $\hat{F}$ and $\hat{F}^\dag$
with those in the normal state, $\hat{G}^0$ and $\hat{F}=\hat{F}^\dag=0$.
We denote the eigenvalue as $\lambda''$.
Figure \ref{fig:fig5} shows the $T$-dependence of $\lambda'$ and $\lambda''$.
We see that $\lambda'$ is strongly suppressed,
and it does not reach unity.
On the other hand, $\lambda''$ is almost equal to $\lambda$ and exceed unity at $T\lesssim T_{cp}$.
Therefore, the drastic increase in $\lambda$ under $T_{cd}$ is mainly due to the ABS-driven FM fluctuations.

\subsection{Result of the $GV^I$-FLEX approximation}
In this study,
we analyze the linearized triplet gap equation
for the pairing interaction calculated by the $GV^I$-FLEX approximation
in the ($1,1$) edge cluster model \cite{matsubara_abs_fm}.
In the conventional FLEX,
the negative feedback effect on spin susceptibility near an impurity is overestimated
since the vertex corrections for the spin susceptibility is not considered \cite{Kontani-imp}.
In the modified FLEX,
the cancellation between negative feedback and vertex corrections
is assumed, and then reliable results are obtained for the single impurity problem \cite{Kontani-imp}.

${\Delta_0}^*$ is the renormalized gap by the normal self-energy.
We obtain ${\Delta_0}^*\approx0.087$ and $2{\Delta_{\rm max}}^*/T_{cd}\approx7.69$ for $\Delta_0=0.12$,
and ${\Delta_0}^*\approx0.058$ and $2{\Delta_{\rm max}}^*/T_{cd}\approx5.11$ for $\Delta_0=0.08$.
To simplify the analysis, the normal self-energy is not included in the Green functions in the gap equation.

\begin{figure}[h]
\includegraphics[width=0.7\linewidth]{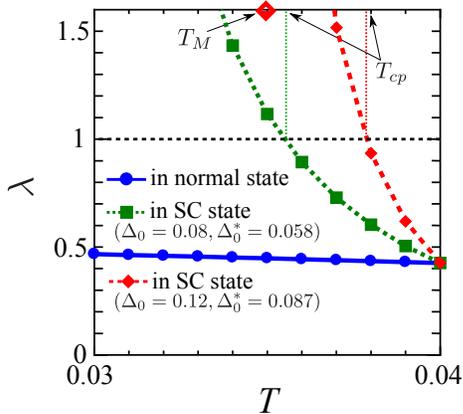}
\caption{(color online)
$T$-dependence of $\lambda$ for the pairing interaction by the $GV^I$-FLEX.
${\Delta_0}^*$ is renormalized gap by the self-energy.
We obtain ${\Delta_0}^*=0.058$ for ${\Delta_0}=0.08$ and ${\Delta_0}^*=0.087$ for ${\Delta_0}=0.12$.
}
\label{fig:fig6}
\end{figure}

Figure \ref{fig:fig6} shows the $T$-dependence of $\lambda$ based on the $GV^I$-FLEX.
$\lambda$ increases as $T$ decreases also in the $GV^I$-FLEX.
In the case of $\Delta_0=0.08$, $\lambda$ exceeds unity at $T\approx0.02$.
For $\Delta_0=0.12$, the increase in $\lambda$ is sharper than that for $\Delta_0=0.08$
because of the stronger development of the FM fluctuations.
The increase in $\lambda$ becomes milder than that in the RPA due to the negative feedback effect of self-energy.
However, we obtain the emergence of a $d \pm ip$-wave superconductivity even if the self-energy is considered.
Note that the $T$-dependence of $\lambda$ based on the RPA and $GV^I$-FLEX is comparable
when
$(2{\Delta_{\rm max}}/T_{cd})_{\rm RPA}\approx(2{\Delta_{\rm max}}^*/T_{cd})_{\rm FLEX}$.

\subsection{Effect of finite $d$-wave coherence length on edge-induced triplet superconductivity}
In this section, we discuss the emergence of the $p$-wave superconductivity
when the $d$-wave gap is suppressed for the finite range $1\leq y \leq \xi_d$,
where $\xi_d$ is the coherence length of the $d$-wave superconductivity.
We set the $y$-dependence of the $d$-wave gap as follows:
\begin{align}
\Delta_{y,y'}(k_x,T)
\left(1-\exp\left(\frac{y+y'-2}{2\xi_d}\right)\right).
\label{eq:gap-xd}
\end{align}
We note that the SC FLEX approximation \cite{Takimoto_FLEX} is applied to the edge cluster model,
the obtained $d$-wave gap for $y\lesssim \xi_d$ should be naturally suppressed.
Instead, we set $\xi_d$ as a parameter to simplify the analysis.
From the experimental results
\cite{cuprate_coherence_3,cuprate_coherence_4,cuprate_coherence_5,cuprate_lattice_1},
we can estimate $\xi_d$ to be 3 sites for $T\ll T_{cd}$.
For $T\lesssim T_{cd}$, 
$\xi_d\gg 3$ because of the relation $\xi_d \propto (1-T/T_{cd} )^{-1/2}$ in the GL theory.
Thus, we set $\xi_d=3$ and 10 in the present analysis.

\begin{figure}[h]
\includegraphics[width=0.95\linewidth]{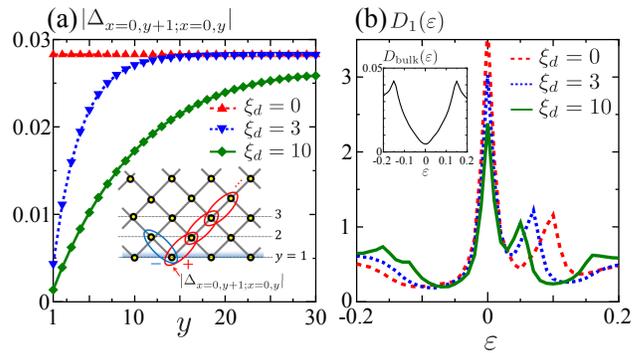}
\caption{(color online)
(a) Site-dependence of $d$-wave gap suppressed near the edge over $\xi_d$.
The inset shows the nearest neighbor bonds corresponding to $|\Delta_{x=0,y+1;x=0,y}|$.
We set $\Delta_{0}=0.08$ and calculated at $T=0.032$.
(b) $\varepsilon$-dependence of LDOS at the ($1,1$) edge for the $d$-wave gap with finite $\xi_d$.
The inset shows the LDOS in the bulk ($y=400$).
}
\label{fig:fig7}
\end{figure}

Figure \ref{fig:fig7} (a) shows the site-dependence of the $d$-wave gap expressed by \eqref{eq:gap-xd}.
Fig. \ref{fig:fig7} (b) shows the obtained LDOS.
At the ($1,1$) edge,
the LDOS has a large peak at $\varepsilon=0$ due to the ABS.
Although the height of the peak becomes lower,
the peak structure due to the ABS still exists for finite $\xi_d$.
The inset is the LDOS in the bulk,
and it shows $V$-shaped $\varepsilon$-dependence
since the $d$-wave gap has line nodes.
In our previous paper, we confirmed that $\alpha_S$ increases as $T$ decreases for finite $\xi_d$.

\begin{figure}[h]
\includegraphics[width=0.95\linewidth]{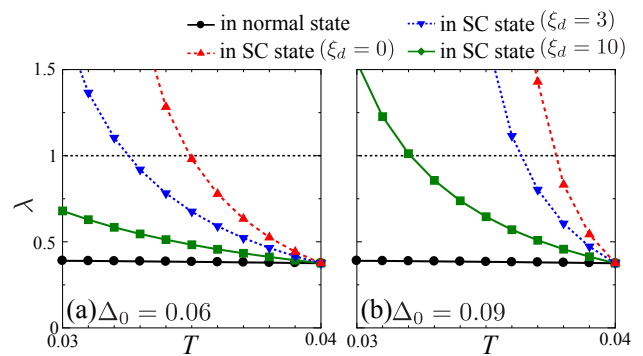}
\caption{(color online)
$T$-dependence of $\lambda$ for (a) $\Delta_{0}=0.06$ or (b) $\Delta_{0}=0.09$ with finite $\xi_d$.
The pairing interaction is calculated by the RPA for finite $\xi_d$.
}
\label{fig:fig8}
\end{figure}

Then, we analyze the gap equation based on the RPA for finite $\xi_d$.
Figure \ref{fig:fig8} shows the $T$-dependence of $\lambda$.
For $\Delta_0=0.09$, 
$\lambda$ increases as the temperature decreases and exceeds unity even for finite $\xi_d$.
On the other hand,
the increase in $\lambda$ is mild for $\Delta_0=0.06$ and $\xi_d$,
and $\lambda \approx 0.68$ even at $T=0.03$. 
Therefore, the strong increase in $\lambda$ is realized under the conditions
$2\Delta_{\rm max}/T_{cd} \gtrsim 6$ and $\xi_d \ll 10$.
These conditions are satisfied in real cuprate superconductors.

\section{cancellation of edge supercurrent in $d \pm ip$-wave SC state}
\label{sec:spincurrent}
In the time-reversal braking SC state,
there is a possibility of the emergence of the edge supercurrent.
In this section, we calculate the edge supercurrent in the $d \pm ip$-wave SC state.
The current operator for $\sigma$-spin electron along $x$-direction
is given as \cite{current_operator}
\begin{eqnarray}
J_{y,y'}^x(k_x)
=
\frac{\partial}{\partial k_x}H^0_{y,y'}(k_x).
\label{eq:current_op}
\end{eqnarray}
Note that $J_{y,y'}^x(k_x)$ does not include the SC gaps.
The spontaneous super current between layer $y$ and layer $y'$ is 
\begin{eqnarray}
\langle J_{y,y'}^x \rangle
=
-
\frac{e}{2}
\sum_{k_x}
\left\{
J_{y,y'}^x(k_x)n_{y,y'}^{\sigma\sigma}(k_x)
+
(y \leftrightarrow y')
\right\},
\label{eq:charge_current}
\end{eqnarray}
where $n_{y,y}^{\sigma\sigma} (k_x)$  is given as
\begin{eqnarray}
n^{\sigma\alpha}_{y,y'}(k_x)
&=&
\langle
c_{k_x,y,\sigma}^\dag c_{k_x,y',\alpha}
\rangle
\nonumber\\
&=&
\sum_{b}
U_{(y\sigma),b}(k_x)U^*_{(y'\alpha),b}(k_x)
\nonumber\\
&\times&
\left\{
T\sum_n
{\rm{Re}}
G_{b}(k_x,\e_n)
+\frac{1}{2}
\right\}.
\label{eq:particle_denity}
\end{eqnarray}
$\hat{U}$ is the unitary matrix to diagonalize BdG hamiltonian in the $d \pm ip$-wave SC state
and $G_{b}$ is Green function in the band representation.
We explain the Green function in the $d \pm ip$-wave SC state in Appendix A.
Here, we define the edge current though the layer $y$ as
\begin{eqnarray}
\langle
J_{y}^x
\rangle
=
\sum_{y'}
\langle J_{y,y'}^x \rangle.
\label{eq:charge_current_2}
\end{eqnarray}
Then, the total super current is given by $\langle J^x \rangle=\sum_{y} \langle J_{y}^x\rangle$. 

\begin{figure}[h]
\includegraphics[width=0.70\linewidth]{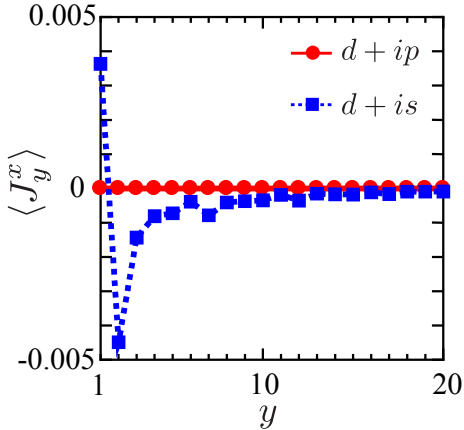}
\caption{(color online)
$y$-dependence of edge supercurrent $\langle J_{y}^x \rangle$
in the $d+ip$- and $d+is$-wave SC state.
We set $\Delta_0=0.09$ and $\underset{i,j}{\rm max}|\phi_{i,j}|=0.09$.
We set the size of edge $s$-wave gap as $\Delta^s=0.09$.
}
\label{fig:fig9}
\end{figure}

Figure \ref{fig:fig9} shows the obtained $y$-dependence of the edge current
in the $d+ip$- and $d+is$-wave SC state.
We set the edge $s$-wave gap as $i\Delta^s\delta_{y,y'=1}$ and $\Delta^s=0.09$ for simplicity.
In the $d+ip$-wave SC state, the time-reversal symmetry is broken.
Nonetheless no edge current does flows.
On the other hand,
the current flows along the edge in the $d+is$-wave SC state
as pointed out in Refs. \cite{Matsumoto-Shiba-II}.

To explain why the spontaneous edge current cancels in the $d+ip$-wave SC state,
we consider the Green function $G_{y,y'}^{\uparrow\uparrow}(k_x,\varepsilon_n)$,
which corresponds to the transfer process of up spin electron from site $y'$ to $y$.
Here, we evaluate an example of its second order term in proportion to $\Delta\phi^\dag$:
\begin{align}
&\delta G_{y,y'}^{\uparrow\uparrow}(k_x,\varepsilon_n)
\nonumber\\
=&
-G_{y,y_1}^0(k_x,\varepsilon_n)
\Delta_{y_1,y_2}^{\uparrow\downarrow}(k_x)
\nonumber\\
&\times
G_{y_3,y_2}^0(-k_x,-\varepsilon_n)
{\phi_{y_3,y_4}^{\uparrow\downarrow}}^{\dag}(k_x)
G_{y_4,y'}^0(k_x,\varepsilon_n),
\label{eq:current_operator_real_space_1}
\end{align}
where $G_{y,y'}^0(k_x,\varepsilon_n)$ is the Green function in the normal state.
Then,
the inverse transfer process of \eqref{eq:current_operator_real_space_1}
contributing to $G_{y',y}^{\uparrow\uparrow}(-k_x,\varepsilon_n)$ is given by
\begin{align}
&\delta G_{y',y}^{\uparrow\uparrow}(-k_x,\varepsilon_n)
\nonumber\\
=&
-G_{y',y_4}^0(-k_x,\varepsilon_n)
\phi_{y_4,y_3}^{\uparrow\downarrow}(-k_x)
\nonumber\\
&\times
G_{y_2,y_3}^0(k_x,-\varepsilon_n)
{\Delta_{y_2,y_1}^{\uparrow\downarrow}}^{\dag}(-k_x)
G_{y_1,y}^0(-k_x,\varepsilon_n).
\label{eq:current_operator_real_space_2}
\end{align}
Note that $\hat{G}^0$ satisfies $G_{y,y'}^0(k_x,\varepsilon_n)=G_{y',y}^0(-k_x,\varepsilon_n)$.
In addition, by using
\eqref{eq:singlet_even-parity}, \eqref{eq:singlet_delta_deltadag},
\eqref{eq:triplet_odd-parity}, and \eqref{eq:condition_obtained-p-wave},
we obtain $\delta G_{y,y'}^{\uparrow\uparrow}(k_x,\varepsilon_n)=\delta G_{y',y}^{\uparrow\uparrow}(-k_x,\varepsilon_n)$.
Therefore, $n^{\sigma\sigma}_{y,y'}(k_x)= n^{\sigma\sigma}_{y',y}(-k_x)$ holds
and therefore the current does not flow.

\section{Summary}
In this paper, we demonstrated that the $d \pm ip$-wave SC state is realized
at the ($1,1$) edge of the $d$-wave superconductors
due to the ABS-induced strong FM fluctuations.
We studied the two-dimensional cluster Hubbard model
with the edge in the presence of the bulk $d$-wave SC gap.
To analyze the edge-induced SC gap,
we constructed a linearized triplet SC gap equation in the presence of the bulk $d$-wave SC gap.
The site-dependent pairing interaction is calculated using the RPA or $GV^I$-FLEX.
The obtained phase difference between the bulk $d$-wave gap and the edge $p$-wave gap is $\pi/2$ in the $\bm k$-space,
and it is the $d \pm ip$-wave SC state
in which the time-reversal symmetry is broken.
Next, we examined the $T$-dependence of the eigenvalue $\lambda$ for the edge-induced SC state.
Below the bulk $d$-wave transition temperature $T_{cd}$,
$\lambda$ for the triplet state increases drastically as $T$ decreases, and exceeds unity at $T=T_{cp}$.
Therefore, the $d \pm ip$-wave SC state is realized at $T_{cp}\lesssim T_{cd}$.
In the $d \pm ip$-wave SC state, the edge current does not flow
irrespective of the time-reversal symmetry braking.

We expect that the $d \pm ip$-wave SC state is also realized
when the direction of the edge is near the ($1,1$) edge
because of the following reason:
The present edge $p$-wave SC is mediated by the ABS-induced strong FM fluctuations,
and the formation of the ABS is confirmed for other edges by the numerical calculations
\cite{Tanaka-ZBCP,Matsumoto-Shiba-III,Tanuma_dpis}.
For the small deviation from the (1,1) edge,
the FM fluctuations should develop and the emergence of
the $d \pm ip$-wave SC state is expected.

The uniqueness of the linearlized edge gap equation \eqref{eq:trip_gap_eq3} is
that only the edge-induced gap is linearized while the effect of the bulk SC gap
is included unperturbatively.
This equation is very useful in analyzing interesting edge-induced superconductivity
in bulk superconductors.
Interesting $d \pm ip$-wave state is naturally obtained
owing to the interference between the bulk and edge gap functions. 

In the present study,
the edge layer can be regarded as the 1-dimensional p-wave superconductor
since the d-wave gap vanishes in the edge layer.
In Ref. \cite{Majorana}, the emergence of the Majorana fermion
at the endpoint of the 1-dimensional p-wave superconductor is proposed.
Therefore, the formation of the Majorana fermion is expected
at the endpoint of the ($1,1$) edge.
Thus, the present study of the edge-induced novel superconductivity
induced by the ABS-driven strong correlation may offer
an interesting platform of SC devises.
Finally, we note that the emergence of the $p$-wave SC and Majorana edge state
had been discussed at the interface between the bulk $s$-wave superconductor and magnetic material
\cite{Majorana_2,Majorana_3}.

\acknowledgements
We are grateful to S. Onari, and Y. Yamakawa for valuable comments and discussions.
This work was supported by the JSPS KAKENHI (No. JP19H05825, No. JP18H01175, and No. JP19J21693).

\appendix
\section{Nambu representation for coexisting SC state in $(k_x,y,y')$-representation}
In this appendix, we explain the Nambu representation in $(k_x,y,y')$-representation.
We assume that the bulk $d$-wave gap
$\Delta_{y,y'}(k_x)\equiv\Delta^{\uparrow\downarrow}_{y,y'}(k_x)$
defined in \eqref{eqn:d-gap}
and the edge triplet gap
$\phi_{y,y'}(k_x)\equiv\phi^{\uparrow\downarrow}_{y,y'}(k_x)$
are both finite.
First, we consider following hamiltonian.
\begin{align}
H
&=
\sum_{k,y,y',\sigma}
H^0_{y,y'}(k_x)
c_{k_x,y,\sigma}^{\dag}
c_{k_x,y,\sigma}
\nonumber\\
&+
\frac{1}{2}
\sum_{k_x,y,y',\sigma\rho}
\left\{
D^{\sigma\rho}_{y,y'}(k_x)
c_{k_x,y,\sigma}^{\dag}
c_{-k_x,y',\rho}^{\dag}
+h.c.
\right\},
\label{eq:BCS_hamiltonian}
\end{align}
where $D_{y,y'}^{\sigma\rho}(k_x)$ is the total gap function,
which includes both singlet $d$-wave gap and triplet gap.
$\sigma$ and $\rho$ represent the spin index.
In this study, we ignore the spin orbit interaction,
so we can set
the d-vector as $\hat{{\bm d}}(k_x)=(0,0,\hat{\phi}(k_x))$,
where hat means $N_y \times N_y$ matrix of sites.
Then, the total gap is given by
\begin{align}
\hat{D}(k_x)
&=
i\hat{d}_0(k_x)\sigma_2+i\hat{{\bm d}}(k_x)\cdot{\bm \sigma}\sigma_2
\nonumber\\
&=
\left(
    \begin{array}{cc}
0
&\hat{\Delta}(k_x)+\hat{\phi}(k_x) \\
-\hat{\Delta}(k_x)+\hat{\phi}(k_x)
&0\\
    \end{array}
\right),
\label{eq:delta_phi}
\end{align}
where ${\bm \sigma}=(\sigma_1,\sigma_2,\sigma_3)$ is the pauli matrix for spin space.
Then,
we obtain the $2N_y \times 2N_y$ Nambu representation as follows:
\begin{align}
H
&=
\sum_{k_x}
\left(
^{t}\hat{c}_{k_x,\uparrow}^{\dag},
^{t}\hat{c}_{-k_x,\downarrow}
\right)
\left(
    \begin{array}{cc}
\hat{H}^{0}(k_x)                                   &     \hat{D}^{\uparrow\downarrow}(k_x) \\
\left\{{\hat{D}^{\uparrow\downarrow}}(k_x)\right\}^\dag    &      -\hat{H}^{0}(-k_x)\\
    \end{array}
\right)
\nonumber\\
&\quad\quad\times
\left(
    \begin{array}{c}
      \hat{c}_{k_x,\uparrow} \\
      \hat{c}_{-k_x,\downarrow}^{\dag} \\
    \end{array}
  \right),
\label{eq:nanbu_2}
\end{align}
where $\hat{c}_{k_x,\uparrow}$ and $\hat{c}_{-k_x,\downarrow}^{\dag}$ represent
the $N_y$-component column vector of sites.
The corresponding Nambu Green function is given as
\begin{align}
&
\left(
    \begin{array}{cc}
      \hat{\mathcal{G}}^{\uparrow\uparrow}(k_x,\varepsilon_n)
&     \hat{\mathcal{F}}^{\uparrow\downarrow}(k_x,\varepsilon_n) \\
      \hat{\mathcal{F}^\dag}^{\uparrow\downarrow}(k_x,\varepsilon_n)
&      -{^{t}\hat{\mathcal{G}}^{\downarrow\downarrow}(-k_x,-\varepsilon_n)}\\
    \end{array}
\right)
\nonumber\\
=&
\left(
    \begin{array}{cc}
      i\varepsilon_n-\hat{H}^{0}(k_x) 
&     -\hat{D}^{\uparrow\downarrow}(k_x) \\
     -\left\{\hat{D}^{\uparrow\downarrow}(k_x)\right\}^\dag
&      i\varepsilon_n+{^{t}\hat{H}^{0}}(-k_x)\\
    \end{array}
\right)^{-1}.
\label{eq:green_all}
\end{align}
$\hat{\mathcal{G}}$, $\hat{\mathcal{F}}$, and $\hat{\mathcal{F}^\dag}$
are the $N_y \times N_y$ Green function in the coexisting SC state.
The Green function in the band representation $G_{b}$
in section IV is obtained by using the superconducting gap equation is expressed as
unitary matrix $\hat{U}$ on \eqref{eq:green_all}.
The 
In this study, we do not consider the frequency dependence of the gap function.
Then, the total gap is represented by the anomalous Green function
as follows:
\begin{align}
D^{\uparrow\downarrow}_{y,y'}(k_x,\varepsilon_n)
=&
T
\sum_{k_x',n',\sigma}
V^{\uparrow\downarrow\sigma\bar{\sigma}}_{y,y'}(k_x-k_x',\varepsilon_n-\varepsilon_n')
\nonumber\\
&\quad\quad\quad
\times
\mathcal{F}^{\sigma\bar{\sigma}}_{y,y'}(k_x',\varepsilon_n'),
\label{eq:gap_anomalous}
\end{align}
where $V^{\rm triplet}_{y,y'}(q_x,i\omega_n)$ is the pairing interaction.
$\bar{\sigma}$ represents the opposite spin to $\sigma$.
In the analysis in the main text,
we do not consider the frequency dependence of the gap function.

\section{Derivation of the linearized triplet gap equation}
In this appendix, we derive
the linearized triplet gap equation
in the presence of the bulk $d$-wave gap.
First, we extract the triplet component
$\phi_{y,y'}(k_x)$
from \eqref{eq:gap_anomalous} by considering
the relation
$
\phi_{y,y'}(k_x)
=
\{
D_{y,y'}^{\uparrow\downarrow}(k_x)
+
D_{y,y'}^{\downarrow\uparrow}(k_x)
\}/2
$
Then, we obtain the equation for the triplet gap
$\phi_{y,y'}(k_x)$ as follows:
\begin{align}
\phi_{y,y'}(k_x)
=
T
\sum_{k_x',n}
&
V^{\rm triplet}_{y,y'}(k_x-k_x',\varepsilon_n-\varepsilon_0)
\nonumber\\
\times
&F_{y,y'}^{\rm triplet}(k_x',\varepsilon_n)
\label{eq:gap_eq_derivation_2}
\end{align}
where
$
F^{\rm triplet}_{y,y'}(k_x,\varepsilon_n)
\equiv
\{
\mathcal{F}^{\uparrow\downarrow}_{y,y'}(k_x,\varepsilon_n)
+\mathcal{F}^{\downarrow\uparrow}_{y,y'}(k_x,\varepsilon_n)
\}/2
$
is triplet part of anomalous Green function in the coexisting SC state.
$
V^{\rm triplet}_{y,y'}(q_x,i\omega_n)
\equiv
{V}^{\uparrow\downarrow\uparrow\downarrow}_{y,y'}(q_x,i\omega_n)
+
{V}^{\uparrow\downarrow\downarrow\uparrow}_{y,y'}(q_x,i\omega_n)
$
is the pairing interaction for triplet SC,
which corresponds to \eqref{eq:trip_gap_int}.
Here, we derive the linearized triplet gap equation
in the presence of finite $d$-wave gap from \eqref{eq:gap_eq_derivation_2}.
For this purpose, we expand the full Nambu Green function in \eqref{eq:green_all}
with respect to $\hat{\phi}$ and $\hat{\phi}^\dag$,
using the following identity:
\begin{align}
\displaystyle
\eqref{eq:green_all}
=&
\left\{
\left(
    \begin{array}{cc}
      \hat{G} & \hat{F} \\
      \hat{F}^\dag & -\hat{\bar{G}} \\
    \end{array}
\right)^{-1}
-
\left(
    \begin{array}{cc}
      0 & \hat{\phi} \\
      \hat{\phi}^\dag &0 \\
    \end{array}
\right)
\right\}^{-1}
\nonumber\\
=&
\left(
    \begin{array}{cc}
      \hat{G} & \hat{F} \\
      \hat{F}^\dag & -\hat{\bar{G}} \\
    \end{array}
\right)
\nonumber\\
+&
\left(
    \begin{array}{cc}
      \hat{G}
      \hat{\phi}
      \hat{F}^{\dag}
    +\hat{F}
      \hat{\phi}^{\dag}
      \hat{G}
&   -\hat{G}
      \hat{\phi}
	\hat{\bar{G}}
    +\hat{F}
      \hat{\phi}^{\dag}
      \hat{F}
\nonumber\\
      \hat{F}^\dag
      \hat{\phi}
      \hat{F}^\dag
	-\hat{\bar{G}}
      \hat{\phi}^\dag
      \hat{G}
&   -\hat{F}^{\dag}
      \hat{\phi}
	\hat{\bar{G}}
	-\hat{\bar{G}}
      \hat{\phi}^{\dag}
      \hat{F}
\\
    \end{array}
\right)
\nonumber\\
+&\ {\rm higher\ order\ terms\ of}\ \phi\ {\rm and}\ \phi^\dag.
\label{eq:green_expand}
\end{align}
where 
$\hat{G}\equiv{\hat{G}}(k_x,\varepsilon_n)$,
$\hat{F}\equiv{\hat{F}}(k_x,\varepsilon_n)$,
$\hat{F}^\dag\equiv{\hat{F}}^\dag(k_x,\varepsilon_n)$,
$\hat{\bar{G}}\equiv{^{t}\hat{G}}(-k_x,-\varepsilon_n)$
are the Green function in the pure $d$-wave SC state
introduced in \eqref{eq:sc-4} in the main text.
The second term in the right-hand-side of \eqref{eq:green_expand}
is the first order terms of $\hat{\phi}$ and $\hat{\phi}^\dag$.
Since $\hat{F}$ satisfies the relation in \eqref{eq:singlet-f},
we obtain the relation
$
\hat{F}^{\rm triplet}
=
   -\hat{G}
      \hat{\phi}
	\hat{\bar{G}}
    +\hat{F}
      \hat{\phi}^{\dag}
      \hat{F}
$.
By substituting it into \eqref{eq:gap_eq_derivation_2},
we obtain the linearized triplet gap equation
in the presence of bulk $d$-wave gap, equation \eqref{eq:trip_gap_eq1}.
We obtain the equation \eqref{eq:trip_gap_eq2} in the same way.
The triplet gap becomes finite
when the eigenvalue $\lambda$ in eqs. \eqref{eq:trip_gap_eq1} and \eqref{eq:trip_gap_eq2} reaches unity.

\section{LDOS in the $d \pm ip$-wave SC state}
Here, we discuss the LDOS in the $d+ip$-wave SC state. 
We assume that the d-vector of the $p$-wave superconductivity is normal to $xy$ plane.
We use the $p$-wave gap obtained by the numerical analysis.
The LDOS is given by
\begin{align}
\displaystyle D_y(\e)=
\frac1{\pi}
\sum_{k_x,\sigma}
{\rm Im} \mathcal{G}^{\sigma,\sigma}_{y,y}(k_x,\e-i\delta).
\end{align}
We set $\delta=0.01$ in the numerical calculation.
\begin{figure}[h]
\includegraphics[width=0.7\linewidth]{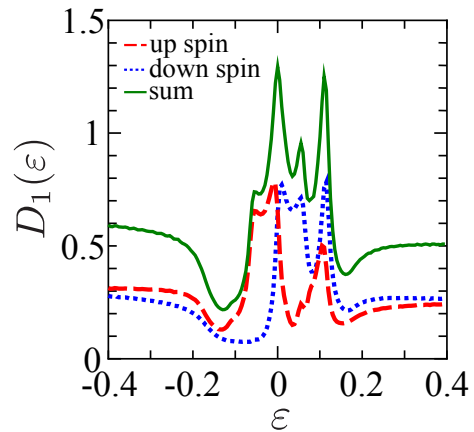}
\caption{(color online)
$\varepsilon$-dependence of the LDOS at the ($1,1$) edge in the $d+ip$-wave SC state.
We set $\Delta_0=0.09$
and $\underset{i,j}{\rm max}|\phi_{i,j}|=0.05$.
The red dashed line and blue doted line represent the LDOS for up and down spin, respectively.
The green solid line is the sum of spins.
}
\label{fig:dpip_ldos}
\end{figure}
Figure \ref{fig:dpip_ldos} shows the obtained LDOS at the edge.
The LDOS for up spin electrons and that for down spin electrons are separated
since the time-reversal symmetry is broken in the $d\pm ip$-wave SC state.



\begin{thebibliography}{99}

\makeatletter
\providecommand \@ifxundefined [1]{%
 \@ifx{#1\undefined}
}%
\providecommand \@ifnum [1]{%
 \ifnum #1\expandafter \@firstoftwo
 \else \expandafter \@secondoftwo
 \fi
}%
\providecommand \@ifx [1]{%
 \ifx #1\expandafter \@firstoftwo
 \else \expandafter \@secondoftwo
 \fi
}%
\providecommand \natexlab [1]{#1}%
\providecommand \enquote  [1]{``#1''}%
\providecommand \bibnamefont  [1]{#1}%
\providecommand \bibfnamefont [1]{#1}%
\providecommand \citenamefont [1]{#1}%
\providecommand \href@noop [0]{\@secondoftwo}%
\providecommand \href [0]{\begingroup \@sanitize@url \@href}%
\providecommand \@href[1]{\@@startlink{#1}\@@href}%
\providecommand \@@href[1]{\endgroup#1\@@endlink}%
\providecommand \@sanitize@url [0]{\catcode `\\12\catcode `\$12\catcode
  `\&12\catcode `\#12\catcode `\^12\catcode `\_12\catcode `\%12\relax}%
\providecommand \@@startlink[1]{}%
\providecommand \@@endlink[0]{}%
\providecommand \url  [0]{\begingroup\@sanitize@url \@url }%
\providecommand \@url [1]{\endgroup\@href {#1}{\urlprefix }}%
\providecommand \urlprefix  [0]{URL }%
\providecommand \Eprint [0]{\href }%
\providecommand \doibase [0]{http://dx.doi.org/}%
\providecommand \selectlanguage [0]{\@gobble}%
\providecommand \bibinfo  [0]{\@secondoftwo}%
\providecommand \bibfield  [0]{\@secondoftwo}%
\providecommand \translation [1]{[#1]}%
\providecommand \BibitemOpen [0]{}%
\providecommand \bibitemStop [0]{}%
\providecommand \bibitemNoStop [0]{.\EOS\space}%
\providecommand \EOS [0]{\spacefactor3000\relax}%
\providecommand \BibitemShut  [1]{\csname bibitem#1\endcsname}%
\let\auto@bib@innerbib\@empty


\bibitem{Bickers}  
 N. E. Bickers and S. R. White, Phys. Rev. B {\bf 43} 8044 (1991).

\bibitem{Monthoux_FLEX}
P. Monthoux and D. J. Scalapino, Phys. Rev. Lett. {\bf 72} 1874 (1994).

\bibitem{Koikegami_FLEX}
S. Koikegami, S. Fujimoto and K. Yamada, J. Phy. Soc. Jpn. {\bf 66} 1438 (1997).

\bibitem{Takimoto_FLEX}
T. Takimoto and T. Moriya, J. Phy. Soc. Jpn. {\bf 66} 2459 (1997).

\bibitem{Dahm_FLEX}
T. Dahm, D. Manske and L. Tewordt, Europhys. Lett. {\bf 55} 93 (2001).

\bibitem{Manske_FLEX}
D. Manske, I. Eremin and K.H. Bennemann, Phys. Rev. B {\bf 67} 134520 (2003).


\bibitem{Moriya}
 T. Moriya and K. Ueda: Adv. Phys. {\bf 49}, 555 (2000).

\bibitem{Moriya-2}
 T. Moriya and K. Ueda, Rep. Prog. Phys. {\bf 66}, 1299 (2003).

\bibitem{Pines}
 P. Monthoux and D. Pines, Phys. Rev. B {\bf 47}, 6069 (1993).

\bibitem{Kontani-rev}
H. Kontani, Rep. Prog. Phys. {\bf 71}, 026501 (2008).

\bibitem{Kontani-Hall}
H. Kontani, K. Kanki, and K. Ueda  Phys. Rev. B {\bf 59}, 14723 (1999).

\bibitem{Kontani-thermoelectric}
H. Kontani, J. Phys. Soc.Jpn, {\bf 70}, 2840 (2001);
H. Kontani, Phys. Rev. Lett. {\bf 89}, 237003 (2002).

\bibitem{Kontani-Nernst-magresi}
H. Kontani, Phys. Rev. B. {\bf 64}, 054413 (2001).


\bibitem{CDW_Ghiringhelli}
G. Ghiringhelli, M. L. Tacon, M. Minola, S. Blanco-Canosa, C.
Mazzoli, N. B. Brookes, G. M. D. Luca, A. Frano, D. G. Hawthorn, F.
He, T. Loew, M. M. Sala, D. C. Peets, M. Salluzzo, E. Schierle, R.
Sutarto, G. A. Sawatzky, E. Weschke, B. Keimer, and L. Braicovich,
Science {\bf 337}, 821 (2012).

\bibitem{CDW_Chang}
J. Chang, E. Blackburn, A. T. Holmes, N. B. Christensen,
J. Larsen, J. Mesot, R. Liang, D. A. Bonn, W. N. Hardy,
A. Watenphul, M. von Zimmermann, E. M. Forgan, and
S. M. Hayden, Nat. Phys. {\bf 8}, 871 (2012).

\bibitem{CDW_Fujita}
K. Fujita, M. H. Hamidian, S. D. Edkins, C. K. Kim, Y. Kohsaka,
M. Azuma, M. Takano, H. Takagi, H. Eisaki, S. Uchida, A. Allais,
M. J. Lawler, E. A. Kim, S. Sachdev, and J. C. Davis, Proc. Natl. Acad.
Sci. U.S.A. {\bf 111}, E3026 (2014).

\bibitem{uniform_CDW_Matsuda}
Y. Sato, S. Kasahara, H. Murayama, Y. Kasahara, E.-G. Moon,
T. Nishizaki, T. Loew, J. Porras, B. Keimer, T. Shibauchi, and
Y. Matsuda, Nat. Phys. {\bf 13}, 1074 (2017).


\bibitem{Chubukov_CDW}
Y. Wang and A. V. Chubukov, Phys. Rev. B {\bf 90}, 035149 (2014).

\bibitem{Kivelson_CDW}
E. Berg, E. Fradkin, S. A. Kivelson, and J. M. Tranquada, New J. Phys. {\bf 11}, 115004 (2009).

\bibitem{Sachdev_CDW}
M. A. Metlitski and S. Sachdev, New J. Phys. {\bf 12}, 105007 (2010);
S. Sachdev and R. La Placa, Phys. Rev. Lett. {\bf 111}, 027202 (2013).

\bibitem{Onari-CDW}
S. Onari, Y. Yamakawa and H. Kontani,
Rev. Lett. {\bf 116}, 227001 (2016).

\bibitem{Yamakawa-CDW}
Y. Yamakawa and H. Kontani, Phys. Rev. Lett. {\bf 114}, 257001 (2015).

\bibitem{Kawaguchi-CDW}
K. Kawaguchi, Y. Yamakawa, M. Tsuchiizu, and H. Kontani,
J. Phys. Soc. Jpn. {\bf 86}, 063707 (2017).

\bibitem{Alloul99-2}
P. Mendels, J. Bobroff, G. Collin, H. Alloul, M. Gabay, 
J. F. Marucco, N. Blanchard and B. Grenier,
Europhys. Lett. {\bf 46}, 678 (1999).

\bibitem{Ishida96}
K. Ishida, Y. Kitaoka, K. Yamazoe, K. Asayama, and Y. Yamada,
Phys. Rev. Lett. {\bf 76}, 531 (1996).

\bibitem{Alloul94}
A. V. Mahajan, H. Alloul, G. Collin, and J. F. Marucco,
Phys. Rev. Lett. {\bf 72}, 3100 (1994).

\bibitem{Alloul00}
W. A. MacFarlane, J. Bobroff, H. Alloul, P. Mendels, N. Blanchard, G. Collin, and J.-F. Marucco,
Phys. Rev. Lett. {\bf 85}, 1108 (2000).

\bibitem{Alloul00-2}
 A. V. Mahajan, H. Alloul, G. Collin, J. F.Marucco,  
 Eur. Phys. J. {\bf B}, 13 457 (2000). 

\bibitem{Alloul99}
J. Bobroff, W. A. MacFarlane, H. Alloul, P. Mendels, N. Blanchard, G. Collin, and J.-F. Marucco, 
Phys. Rev. Lett. {\bf 83}, 4381 (1999).

\bibitem{Bulut00}
N. Bulut, Phys. Rev. B {\bf 61}, 9051 (2000).

\bibitem{Ohashi_imp_RPA}
Y. Ohashi, J. Phys. Soc. Jpn. {\bf 70}, 2054 (2001).

\bibitem{Kontani-imp}
H. Kontani and M. Ohno,
Phys. Rev. B {\bf 74}, 014406 (2006);
H. Kontani and M. Ohno,
J. Magn. Magn. Mat. {\bf 310}, 483 (2007).

\bibitem{Matsubara-edge}
S. Matsubara, Y. Yamakawa, H. Kontani,
J. Phys. Soc. Jpn {\bf 87}, 073705 (2018).

\bibitem{imp_Harter}
J. W. Harter, B. M. Andersen, J. Bobroff, M. Gabay, and P. J. Hirschfeld
Phys. Rev. B {\bf 75}, 054520 (2007).

\bibitem{suf_Andersen}
Brian M. Andersen, Ashot Melikyan, Tamara S. Nunner, and P. J. Hirschfeld
Phys. Rev. Lett. {\bf 96}, 097004 (2006).


\bibitem{matsubara_abs_fm}
S. Matsubara and H. Kontani Phys. Rev. B {\bf 101}, 075114 (2020).


\bibitem{Hu-ZBCP}
C. R. Hu, Phys. Rev. Lett. {\bf 72}, 1526 (1994).

\bibitem{Tanaka-ZBCP}
Y. Tanaka and S. Kashiwaya, Phys. Rev. Lett. {\bf 74}, 3451 (1995).

\bibitem{Kashiwaya-junction}
S. Kashiwaya, Y. Tanaka, M. Koyanagi, K. Kajimura, Phys. Rev. B {\bf 53}, 2667 (1996).

\bibitem{Matsumoto-Shiba-ABS}
M. Matsumoto and H. Shiba, J. Phys. Soc. Jpn. {\bf 64}, 1703 (1995).

\bibitem{Nagato}
Y. Nagato and K. Nagai, Phys. Rev. B {\bf 51}, 16254 (1995).

\bibitem{Kashiwaya-ZBCP}
S. Kashiwaya and Y. Tanaka, Rep. Prog. Phys. {\bf 63}, 1641 (2000).


\bibitem{Kashiwaya-ZBCP-2}
S. Kashiwaya, Y. Tanaka, M. Koyanagi, H. Takashima, and K. Kajimura, Phys. Rev. B {\bf 51}, 1350 (1995).

\bibitem{Iguchi}
I. Iguchi, W. Wang, M. Yamazaki, Y. Tanaka, and S. Kashiwaya, Phys. Rev. B {\bf 62}, R6131 (2000).

\bibitem{Wei-ZBCP}
J. Y. T. Wei, N. -C. Yeh, D. F. Garrigus, and M. Strasik, Phys. Rev. Lett. {\bf 81}, 2542 (1998).

\bibitem{Geek-ZBCP}
J. Geek, X. X. Xi, and G. Linker, Z. Phys. B {\bf 73}, 2542 (1988).


\bibitem{triplet_FM_fluc_Fay}
D. Fay and J. Appel, Phys. Rev. B {\bf 22}, 3173 (1980). 

\bibitem{triplet_FM_fluc_Monthoux}
P. Monthoux and G. G. Lonzarich,Phys. Rev. B {\bf 59},  14598 (1999).

\bibitem{triplet_FM_fluc_Wang}
Z. Wang, W. Mao, and K. Bedell, Phys. Rev. Lett. {\bf 87}, 257001 (2001).

\bibitem{triplet_FM_fluc_Roussev}
R. Roussev and A. J. Millis, Phys. Rev. B {\bf 63}, 140504(R) (2001).

\bibitem{triplet_FM_fluc_Fujimoto}
S. Fujimoto, J. Phys. Soc. Jpn. {\bf 73}, 2061 (2004). 


\bibitem{Matsumoto-Shiba-I}
M. Matsumoto and H. Shiba, J. Phys. Soc. Jpn. {\bf 64}, 3384 (1995).

\bibitem{Matsumoto-Shiba-II}
M. Matsumoto and H. Shiba, J. Phys. Soc. Jpn. {\bf 64}, 4867 (1995).

\bibitem{Matsumoto-Shiba-III}
M. Matsumoto and H. Shiba, J. Phys. Soc. Jpn. {\bf 65}, 2194 (1995).

\bibitem{Tanuma_dpis}
Y. Tanuma, Y. Tanaka, M. Ogata, and S. Kashiwaya, Phys. Rev. B {\bf 60}, 9817 (1999).

\bibitem{zbcp_split_1}
S. Kashiwaya, Y. Tanaka, M. Koyanagi, H. Takashima, and K. Kajimura,
J. Phys. Chem. Solids {\bf 56}, 1721 (1995).

\bibitem{zbcp_split_2}
Y. Tanaka, Y. Tanuma, and S. Kashiwaya,
Phys. Rev. B {\bf 64}, 054510 (2001).

\bibitem{zbcp_split_3}
Y. Tanuma, Y. Tanaka, and S. Kashiwaya,
Phys. Rev. B {\bf 64}, 214519 (2001).

\bibitem{Kuboki-GL_jpsj}
K. Kuboki and M. Sigrist, J. Phys. Soc. Jpn. {\bf 65}, 361 (1995).

\bibitem{Sigrist-Kuboki-TB}
M. Sigrist, K. Kuboki, P. A. Lee, A. J. Millis, T. M. Rice, Phys. Rev. B {\bf 53}, 2835 (1996).

\bibitem{Kuboki-TB_t-J}
K. Kuboki and M. Sigrist, J. Phys. Soc. Jpn. {\bf 67}, 2873 (1998).

\bibitem{Watashige-FeSe-TB}
T. Watashige, Y. Tsutsumi, T. Hanaguri, Y. Kohsaka, S. Kasahara, A. Furusaki, M. Sigrist,
C. Meingast, T. Wolf, H. v. Löhneysen, T. Shibauchi, and Y. Matsuda
Phys. Rev. X {\bf 5}, 031022 (2015).


\bibitem{supercurrent_1}
M. H\r{a}kansson, T. L\"{o}fwander, and M. Fogelstr\"{o}m,
Nat. Phys. {\bf 11}, 755 (2015).

\bibitem{supercurrent_2}
P. Holmvall, A.B. Vorontsov, M. Fogelstr\"{o}m, and T. L\"{o}fwander,
Nat. Commun. {\bf 9}, 2190 (2018).

\bibitem{cuprate_coherence_1}
D. S. Inosov, J. T. Park, A. Charnukha, Yuan Li, A. V. Boris, B. Keimer, and V. Hinkov
Phys. Rev. B {\bf 83}, 214520 (2011).

\bibitem{cuprate_coherence_2}
{\O}ystein Fischer, Martin Kugler, Ivan Maggio-Aprile, Christophe Berthod, and Christoph Renner
Rev. Mod. Phys. {\bf 79}, 353 (2007).

\bibitem{cuprate_coherence_3}
Y. Matsuda, T. Hirai, S. Komiyama, T. Terashima, Y. Bando, K. Iijima, K. Yamamoto, and K. Hirata
Phys. Rev. B {\bf 40}, 5176 (1989).

\bibitem{cuprate_coherence_4}
K. Semba, A. Matsuda, and T. Ishii
Phys. Rev. B {\bf 49}, 10043 (1994).

\bibitem{cuprate_coherence_5}
K. Tomimoto, I. Terasaki, and A. I. Rykov, T. Mimura, and S. Tajima
Phys. Rev. B {\bf 60}, 114 (1999).


\bibitem{cuprate_lattice_1}
F. Izumi, H. Asano, T. Ishigaki, A. Ono, and F. P. Okamura
Jpn. J. Appl. Phys. {\bf 26}, L611 (1987).

\bibitem{current_operator}
A. C. Durst and P. A. Lee
Phys. Rev. B {\bf 62}, 1270 (2000).

\bibitem{Majorana}
A. Y. Kitaev, Phys. Usp. {\bf 44}, 131 (2001).

\bibitem{Majorana_2}
S. Nakosai, Y. Tanaka, and N. Nagaosa,
Phys. Rev. B {\bf 88}, 180503(R) (2013). 

\bibitem{Majorana_3}
W. Chen and A. P. Schnyder
Phys. Rev. B {\bf 92}, 214502 (2015). 

\end{thebibliography}
\end{document}